\documentclass[twocolumn,showpacs,amsmath,amssymb,aps,prl,floatfix,nofootinbib,superscriptaddress]{revtex4}
\usepackage{graphicx,graphics,times,color}
\usepackage{bm}
\usepackage{longtable}
\usepackage{color}

\def\be{\begin{equation}}
\def\ee{\end{equation}}
\def\ba{\begin{eqnarray}}
\def\ea{\end{eqnarray}}
\def\la{\langle}
\def\ra{\rangle}

\begin{document}

\title{ An order parameter for impurity systems at quantum criticality}

\author{Abolfazl Bayat}
\affiliation{Department of Physics and Astronomy, University College
London, Gower St., London WC1E 6BT, United Kingdom}

\author{Henrik Johannesson}
\affiliation{Department of Physics, University of Gothenburg, SE 412 96 Gothenburg, Sweden}

\author{Sougato Bose}
\affiliation{Department of Physics and Astronomy, University College
London, Gower St., London WC1E 6BT, United Kingdom}

\author{Pasquale Sodano}
\affiliation{International Institute of Physics,
Universidade Federal do Rio Grande do Norte,
59078-400 Natal-RN, Brazil,
and \\
Departamento de F«isica Te\'{o}rica e Experimental, «
Universidade Federal do Rio Grande do Norte,
59072-970 Natal-RN, Brazil}
\affiliation{INFN, Sezione di Perugia, Via A. Pascoli, 06123, Perugia, Italy}

\begin{abstract}
A quantum phase transition may occur in the ground state of a system at zero temperature when a controlling field or interaction is varied. The resulting quantum fluctuations which trigger the transition produce scaling behavior of various observables, governed by universal critical exponents. A particularly interesting class of such transitions appear in systems with quantum impurities where a non-extensive term in the free energy becomes singular at the critical point. Curiously, the notion of a conventional order parameter which exhibits scaling at the critical point is generically missing in these systems. We here explore the possibility to use the Schmidt gap, which is an observable obtained from the entanglement spectrum, as an order parameter. A case study of the two-impurity Kondo model confirms that the Schmidt gap faithfully captures the scaling behavior by correctly predicting the critical exponent of the dynamically generated length scale at the critical point.
\end{abstract}

\maketitle

A quantum phase transition (QPT) occurs at zero temperature as one varies some control parameter \cite{Sachdev}. While a QPT cannot be accessed directly in the laboratory, it leaves distinct imprints at finite temperature: In the case of a continuous QPT, characterized by an avoided level crossing in the ground state energy, large quantum fluctuations in the neighborhood of the quantum critical point spawns a ``quantum critical regime'' where the scaling of observables are encoded in non-integer exponents. This is the signature of a non-Fermi liquid \cite{Schofield}, exemplified in the physics of heavy fermions materials \cite{Si}, unconventional superconductors \cite{Shibauchi}, and quantum impurity systems \cite{IA}.

One usually thinks of a QPT as due to large quantum fluctuations in some local order parameter. Loosely speaking, at the quantum critical point the fluctuations kill off the ground state expectation value of  the local operator which defines the order parameter, signaling that the system has entered a ``disordered" phase. This is but a quantum paraphrase of the Landau-Ginzburg-Wilson (LGW) synopsis of a classical second-order phase transition \cite{Amit}. While there are other less common types of quantum criticality, an order parameter can often still be defined. For example, a symmetry-protected topological phase is characterized by the non-vanishing of a nonlocal order parameter \cite{Chen}. At a deconfined quantum critical point, there is instead a direct QPT between two phases of matter which are each characterized by its own unique non-vanishing local order parameter \cite{Senthil,Laughlin}. But what is the order parameter that comes into play when describing an impurity quantum phase transition (iQPT)?

iQPTs occur in systems where the presence of one or more quantum impurities adds a non-extensive term to the free energy which becomes singular at a quantum critical point \cite{VojtaReview}. As a result $-$ when considering fermions $-$ non-Fermi liquid behavior emerges \cite{IA}, independent of any possible bulk phase transitions. However, although iQPTs have been in the spotlight for several decades now,  with the possible exceptions  of the local magnetization in the pseudogap Kondo \cite{IngersentQi} and sub-ohmic spin-boson \cite{VojtaTongBulla} models, no order parameters have yet been identified. One may have expected that the effective locality and one-dimensionality of quantum impurity interactions \cite{AffleckLesHouches}, should make the search simple: As follows from Coleman's theorem \cite{Coleman}, the spontaneous symmetry breaking usually associated with an order parameter is thereby restricted to only discrete symmetries. However, no explicit discrete symmetries are broken in an iQPT. Alternative known scenarios not encompassing a symmetry breaking, like the Kosterlitz-Thouless transitions \cite{KT} (that appear in quantum impurity models that can be mapped onto the Kondo model with anisotropic exchange \cite{VojtaReview}) are generally inapplicable: In the language of the renormalization group \cite{Amit}, a Kosterlitz-Thouless transition is controlled by a semi-stable fixed point, unlike a generic iQPT where the fixed point is unstable. Similarly, states with intrinsic topological order \cite{Wen} and their associated QPTs are not even a possibility for one-dimensional spinful fermions \cite{Chen}. This poses a conundrum: How to conceptualize iQPTs and relate them to other known classes of QPTs? The rapid progress in experiments on engineered nanoscale systems that now enables unprecedented control and study of non-Fermi liquid physics from quantum impurity criticality \cite{Goldhaber} adds to the importance of finding an answer.

Here we take a new inroad and explore the possibility to use a novel kind of order parameter based on properties of the entanglement spectrum, building on an observation by De Chiara {\em et al.} \cite{sanpera-QPT}. These authors, studying the quantum Ising and spin-1 Heisenberg chains, found that the Schmidt gap defined as $\Delta_{\text{S}}\!=\!\lambda_1\!-\! \!\lambda_2$ \, $-$ where $\lambda_1$ and $\lambda_2$ are the two largest eigenvalues of the reduced ground state density matrix constructed from an arbitrary bipartition of the system $-$ exhibits the same critical scaling as the local magnetization and the mass gap. Thus, the Schmidt gap may be employed as an alternative to a conventional order parameter.
Could an analogous construction be used to identify an order parameter for iQPTs?

Prima facie, this may seem unlikely. Considering the system at equilibrium, an iQPT is encoded in the scaling of impurity response functions which are non-extensive. Indeed, conformal field theory predicts a negligible influence on the Schmidt gap from the impurity degrees of freedom. Using the fact that $\lambda_1/\lambda_2 =
\exp(-\kappa/\ln {\lambda_1})$ (with $\kappa$ a constant) \cite{CalabreseLefevre}, one verifies that $\ln \lambda_1 = -S/2$ in the single-copy entanglement scaling limit, with $S$ the block entanglement of the corresponding bipartitioning of the system \cite{EisertCramer,Peschel}. It follows that $\Delta_{\text{S}} =  \exp (-S/2)(1-\exp(-2\kappa/S))$, with the impurity contribution showing up only as a small non-extensive boundary term in $S$ \cite{AffleckReview}. One should recall, however, that  the information contained in an entanglement spectrum  may depend on how one partitions the Hilbert space of the system. In the conformal field theory approach the ``cut'' has to be taken sufficiently far from the impurities  so as to reach the scaling region \cite{CalabreseCardy} where microscopic processes close to the impurity sites become irrelevant. This is different from a translationally invariant bulk critical system where the cut can be taken anywhere and where conformal field theory correctly predicts the closing of the Schmidt gap \cite{Lepori}. As for an iQPT, by going beyond the confines of conformal theory and taking the cut closer to the impurities, could it be that the Schmidt gap takes notice?
To find out, we shall take as test case a spin-onlyÓ version of the two-impurity Kondo model \cite{bayat-TIKM}. Via a finite-size scaling analysis of a
``close cut'' Schmidt gap we are able to extract the critical exponent for the dynamically generated length scale at the quantum critical point. This confirms that the Schmidt gap indeed captures the correct scaling behavior.

\subsection*{RESULTS}
\begin{center}{ \bf{Two-impurity Kondo model}}\end{center}
The two-impurity Kondo model (TIKM) \cite{Jayprakash} is a paradigm for iQPTs. Here two localized spins of magnitude $S=1/2$ are coupled to the spins of conduction electrons by an antiferromagnetic Kondo interaction and to each other via a Ruderman--Kittel-Kasuya-Yosida (RKKY) interaction. When the Kondo interaction dominates, the impurity spins get screened away, while in the opposite limit they form a local singlet. The crossover between the Kondo and RKKY dominated regimes  sharpens into a QPT when each impurity is coupled to its own separate bath of conduction electrons \cite{Zarand}. Theory predicts non-Fermi liquid response for electron transport and impurity contributions to the thermodynamics \cite{ALJ}, with experimental tests expected in the near future \cite{Goldhaber}. In fact, the quantum phase transitions in the TIKM and the closely related \cite{Mitchell,Mitchell2} overscreened multichannel Kondo models \cite{Jones} are arguably the best understood examples of how quantum criticality produces a non-Fermi liquid.  However, being examples of local quantum critical points $-$ with the ``non-Fermi liquidness'' associated with the interaction of localized and extended degrees of freedom $-$ they are not easily fitted into the conventional picture of quantum criticality \cite{Sachdev}. Indeed, most notably has been the lack of an identifiable zero-field order parameter which exhibits scaling at the quantum phase transition.

\begin{figure} \centering
    \includegraphics[width=8cm,height=7cm,angle=0]{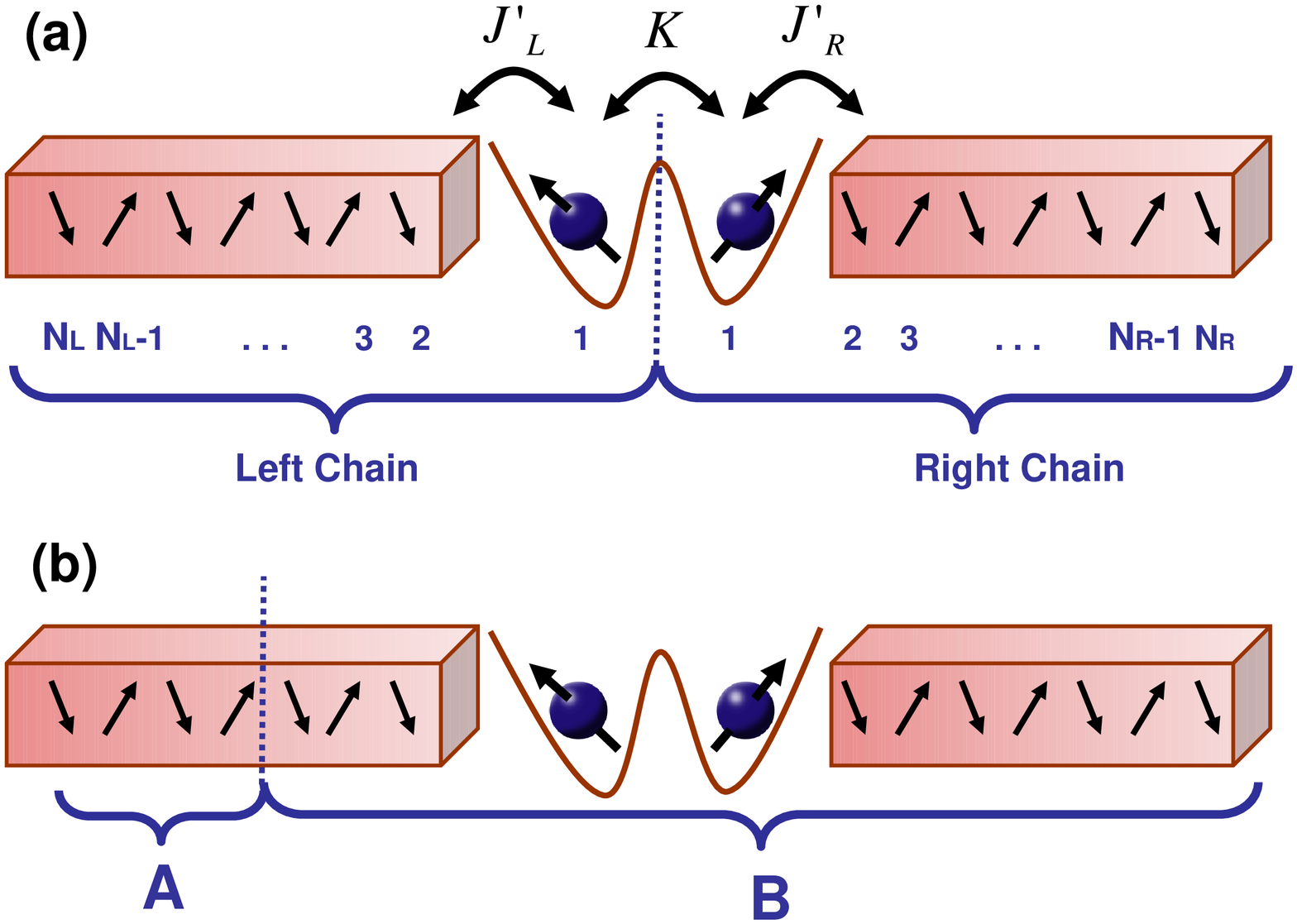}
    \caption{{\bf{Schematic picture of the ``spin-only'' version of the TIKM.}} (a) The two impurities interact with their neighboring spins by Kondo couplings $J'_L$ and $J'_R$ respectively, and with each other via the RKKY coupling $K$; cf. Eq. (\ref{Hamiltonian}). (b) Partitioning of the system into two parts $A$ and $B$, with $n_A$ and $n_B=N_L+N_R-n_A$ spins respectively. The entanglement spectrum with the associated Schmidt gap is obtained by tracing out one of the subsystems.}
         \label{fig1}
\end{figure}

\begin{figure*} \centering
    \includegraphics[width=14cm,height=5cm,angle=0]{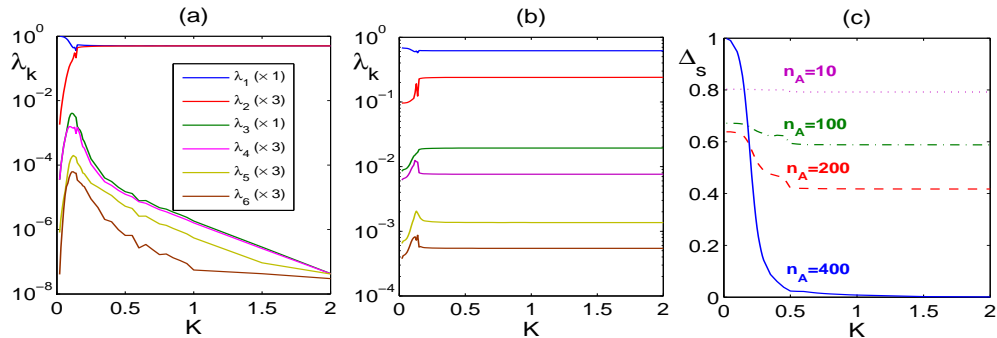}
    \caption{{\bf{Entanglement spectrum and Schmidt gap.}} The first six levels of the entanglement spectrum $\{\lambda_k\}_{k=1}^{6},$ with the Schmidt coefficients $\lambda_k$ as functions of the RKKY coupling $K$ in a chain of $N=800$ and with $J'=0.4$ for (a) symmetric partition with $n_A=n_B=400$; (b) asymmetric partition with $n_A=200$ and $n_B=600$. The number given for each $\lambda_k$ accounts for the corresponding degeneracy. (c) The Schmidt gap versus RKKY coupling $K$ for different partitionings of the system. }
     \label{fig2}
\end{figure*}

To speed up computations, allowing us to obtain high-precision numerical data, we work with a Òspin-onlyÓ version of the TIKM \cite{bayat-TIKM}. Its Hamiltonian is written as $H= \sum_{m=L,R}H_m + H_I$, with
\begin{eqnarray} \label{Hamiltonian}
H_m &\!=\!&  J^{\prime}_m \left( J_1 \boldsymbol{\sigma}_1^m \!\cdot \!\boldsymbol{\sigma}_2^m + J_2 \boldsymbol{\sigma}_1^m \!\cdot \!\boldsymbol{\sigma}_3^m \right) \nonumber \\
&+&
J_1\sum_{i=2}^{N_m-1} \boldsymbol{\sigma}_i^m \!\cdot \!\boldsymbol{\sigma}_{i+1}^m+J_2\sum_{i=2}^{N_m-2} \boldsymbol{\sigma}_i^m \!\cdot \!\boldsymbol{\sigma}_{i+2}^m,
\\
H_I &\!=\! & J_1K \boldsymbol{\sigma}_1^L \!\cdot \!\boldsymbol{\sigma}_1^R. \nonumber
\end{eqnarray}
Here $m=L,R$ labels the left and right chains with $\boldsymbol{\sigma}_i^m$ the vector of Pauli matrices at site $i$ in chain $m$, and with $J_1$  ($J_2$)  nearest- (next-nearest-) neighbor couplings (see Fig.~\ref{fig1}(a)). The parameters $J'_L > 0$ and $J'_R > 0 $ play the role of antiferromagnetic Kondo couplings and $K$ represents the dimensionless RKKY coupling between the impurity spins $\boldsymbol{\sigma}_1^L$ and $\boldsymbol{\sigma}_1^R$. The total size of the system is thus $N=N_L+N_R$. By fine tuning $J_2/J_1$ to the critical point $(J_2/J_1)_c=0.2412$ of the spin chain dimerization transition \cite{Okamoto,Eggert}, the stripped-down version of the TIKM in Eq. (\ref{Hamiltonian}) can be shown to provide a faithful representation of its spin physics \cite{bayat-TIKM} (which is all that matters for describing the iQPT of the model \cite{Zarand,ALJ}).  The procedure is here the same as that employed for the spin-chain emulation of the single-impurity Kondo model \cite{Sorensen}: By tuning $J_2/J_1$ to the dimerization transition all logarithmic scaling corrections vanish, allowing for an unambiguous fit of numerical data.\vspace{3mm}

\begin{center}\bf{Entanglement spectrum and Schmidt gap}\end{center}
Turning to the computation of the Schmidt gap $-$ most efficiently carried out using a Density Matrix Renormalization Group (DMRG) approach$-$ we cut the system in two parts, $A$ and $B$, and write the Schmidt decomposition of the ground state $|GS\ra$ as
\begin{equation}\label{Schmidt-decomp}
    |GS\ra=\sum_{k} \sqrt{\lambda_k} |A_k\ra \otimes |B_k\ra,  \ \ \lambda_k \geq 0,
\end{equation}
with mutually orthogonal Schmidt basis states $|A_k\ra$ and $|B_k\ra$. The density matrix of each part is diagonal in the Schmidt basis,
\begin{equation}\label{rho_LR}
   \rho_\alpha=\sum_{k} \lambda_k |\alpha_k\ra \la \alpha_k|, \ \ \alpha=A,B.
\end{equation}
with the eigenvalues $\lambda_1\geq \lambda_2 \geq ...$ in descending order forming the entanglement spectrum (frequently
defined as $\{- \ln\lambda_i \}_{i=1,2,...}$ in the literature). In Fig.~\ref{fig2}(a) we plot the six largest levels
 as functions of the RKKY coupling $K$ for identical Kondo couplings $J_L' = J_R' = J'$ where the partitioning is obtained by cutting through the bonds joining the two impurities as shown in Fig.~\ref{fig1}(a). Due to the SU(2) symmetry of the system the levels come with different degeneracies, reflecting the singlet-triplet nature of the low-lying eigenstates. One may also make an asymmetric partition (with $A$ and $B$ of different lengths) as shown in Fig.~\ref{fig1}(b). However, when the cut is taken far from the impurities, as in Fig.~\ref{fig2}(b), there is no additional degeneracy in the entanglement spectrum in the RKKY phase, in contrast to the symmetric cut in  Fig.~\ref{fig2}(a).
In Fig.~\ref{fig2}(c) we plot $\Delta_{\text{S}}$ for four different cuts. While the asymmetric cuts far from the impurities yield Schmidt gaps exhibiting a very weak dependence on $K$ (as expected from conformal field theory), the symmetric cut produces a Schmidt gap with a typical finite-size order parameter profile.

\begin{figure} \centering
    \includegraphics[width=9cm,height=7cm,angle=0]{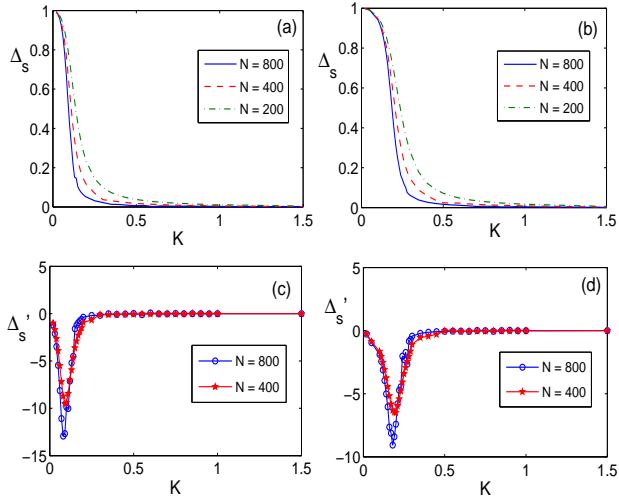}
    \caption{{\bf{The Schmidt gap and its derivative.}} The Schmidt gap $\Delta_{\text{S}}$ versus impurity coupling
     $K$ for (a) $J'=0.4$; (b) $J'=0.5$. The first derivative of the data in the upper panel, i.e. $\Delta_{\text{S}}^{\prime} =\partial \Delta_{\text{S}}/\partial K$ versus $K$ for (c) $J'=0.4$; (d) $J'=0.5$.}
     \label{fig3}
\end{figure}

\begin{figure} \centering
    \includegraphics[width=9cm,height=4cm,angle=0]{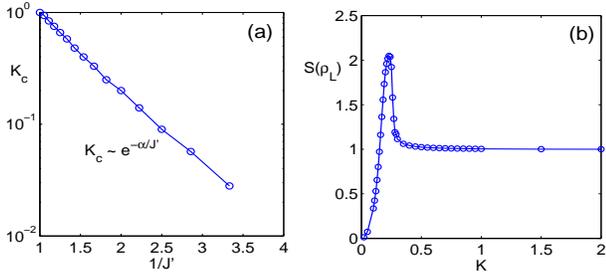}
    \caption{{\bf{Critical coupling and von Neumann entropy.}} (a) The critical coupling $K_c$ as a function of $1/J'$ in a chain of length $N=400$ which shows the exponential scaling predicted in Refs. \cite{JVW,JV}. (b) The von Neumann entropy of $\rho_L$ for the symmetric cut versus the RKKY coupling $K$ for a chain of length $N=800$.}
     \label{fig4}
\end{figure}

To substantiate that the Schmidt gap conveys information about the iQPT, in Figs.~\ref{fig3}(a)-(b) we plot $\Delta_{\text{S}}$ versus $K$ for two different impurity couplings $J^{\prime}$ when the symmetric cut is considered. As revealed by Figs.~\ref{fig3}(a)-(b), the profile becomes sharper for larger chains, suggesting that in the thermodynamic limit, $N\!\rightarrow \!\infty$, the Schmidt gap obtained from a symmetric cut abruptly drops to zero at some critical value, $K=K_c$. We may identify $K_c$ as the point where $\Delta_{\text{S}}^{\prime}=\partial \Delta_{\text{S}}/\partial K$ diverges as $N \!\rightarrow \!\infty$. In Figs.~\ref{fig3}(c)-(d) we plot $\Delta_{\text{S}}^{\prime}$ for the data shown in Figs.~\ref{fig3}(a)-(b) (for clarity of the figures, only two sets of data are shown for each impurity coupling $J^{\prime}$). One sees that the $\Delta_{\text{S}}^{\prime}$ has sharp cusps which become even more pronounced by increasing the system size. This is the finite-size precursor of the non-analyticity in $\Delta_{\text{S}}^{\prime}$ in the thermodynamic limit.  By locating the cusps as functions of the Kondo coupling $J'$, one finds, already for $N\!=\!400$, a near perfect agreement with the known exponential scaling of the TIKM quantum critical point,  $K_c \!\sim\! \exp (-\alpha/J')$ (with $\alpha$ a positive constant) \cite{JVW,JV} as shown in Fig.~\ref{fig4}(a).
For the sake of completeness, in Fig.~\ref{fig4}(b) we also plot the von Neumann entropy $S(\rho)=-Tr(\rho_L\log_2 \rho_L)$ of the left chain (with density matrix $\rho_L$) when the right one is traced out. As expected, for $K=0$, the entropy is zero since the two chains are decoupled and, for $K \gg K_c$, it takes the value $\log_2 2=1$ due to the impurity singlet formation. At the critical point the von Neumann entropy $S(\rho_L)$ takes its maximum value.
\vspace{3mm}

\begin{figure} \centering
    \includegraphics[width=9cm,height=7cm,angle=0]{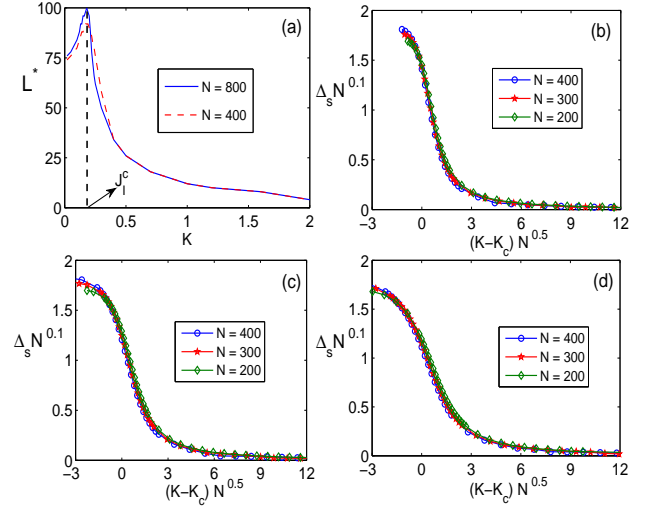}
    \caption{{\bf{Entanglement screening length and finite-size scaling.}}\\ (a) Entanglement screening length $L^*$ as a function of the RKKY coupling $K$ for two different lengths of the chain. The cusp in $L^*$ close to the quantum critical point sharpens as the size of the system increases, reflecting the diverging length scale of the critical system in the thermodynamic limit. (b) Finite size scaling plot of $N^{\beta/\nu}\Delta_{\text{S}}$ vs. $|K-K_c|N^{1/\nu}$ with $\beta=0.2$ and $\nu=2$ for different lengths of the chain, where $J'=0.4$; (c) $J'=0.5$; (d) $J'=0.6$.}
    \label{fig5}
\end{figure}

\begin{center}{\bf{Finite-size scaling}}\end{center}
To further substantiate the interpretation of the Schmidt gap as an order parameter, we have performed a finite-size scaling analysis \cite{Barber}, writing
\begin{equation}\label{FSS}
    \Delta_{\text{S}}=N^{-\beta/\nu} f_{\Delta_{\text{S}}}(|K-K_c|N^{1/\nu}),
\end{equation}
where $f_{\Delta_{\text{S}}}$ is a scaling function, $\beta$ is the critical exponent defined by $\Delta_{\text{S}}\sim |K-K_c|^\beta$, and $\nu$ is the exponent governing the divergence of the crossover scale $\xi$ at the TIKM quantum critical point, $\xi\sim|K-K_c|^{-\nu}$. The corresponding scale $v/\xi \sim |K-K_c|^{\nu}$, with $v$ the velocity of spin excitations, defines the energy below which the theory renormalizes to either the stable (one-impurity) Kondo fixed point ($K\!<\!K_c$) or the local singlet (RKKY) stable fixed point $(K\!>\!K_c)$  \cite{Mitchell, Mitchell3}. Identifying $\xi$ with the Kondo screening length $L^*$ in the neighborhood of the critical point when $K \lesssim K_c$ \cite{ALJ}, one expects the divergence in $\xi$ to reveal itself as a finite-size cusp in $L^*$. Indeed, by defining the screening length as the distance over which the two impurities are maximally entangled with the rest of the system (as quantified by negativity as an entanglement measure \cite{bayat-TIKM}) this is strikingly confirmed in Fig.~\ref{fig5}(a): Here $L^*$ is plotted as a function of $K$ for two different lengths of the chain and one sees that, near the quantum critical point, it shoots up with a peak that becomes sharper for the larger system, signaling a divergence in the thermodynamic limit.

To determine the critical exponents we identify the values of $\beta$ and $\nu$ for which the plot of $N^{\beta/\nu}\Delta_{\text{S}}$ as a function of $|K-K_c|N^{1/\nu}$ collapses to a single curve for chains of different lengths and with different impurity couplings $J'$.  We thus obtain $\beta=0.2 \pm 0.05$ and $\nu=2 \pm 0.1$. This yields an excellent data collapse, as seen in Figs.~\ref{fig5}(b)-(d). The value $\nu\!=\!2$ is in perfect agreement with the results from conformal field theory \cite{ALJ}, corroborating that the Schmidt gap behaves as a proper order parameter.

It is important to realize that the vanishing of the Schmidt gap is a highly nontrivial fact. While it is easy to see that it must vanish deep in the RKKY regime when the two impurities have formed a local singlet, its vanishing, already at intermediate values of $K$ is an intricate many-body effect. Indeed, a direct DMRG computation of the concurrence $C$ \cite{Wootters} between the two impurities reveals that the Schmidt gap closes already when $C\approx 0.7$. In contrast, the fact that the Schmidt gap approaches unity deep in the Kondo regime when $K \rightarrow 0$ is trivially due to the decoupling of the left and right parts of the system: In this limit
$\rho_L$ and $\rho_R$ are pure states resulting in $\lambda_k=\delta_{1k}$ (for $k=1,2,...$), implying that $\Delta_{\text{S}}=1$. \vspace{3mm}

\begin{center}{\bf{Schmidt gap as an observable}}\end{center}
While the Schmidt gap is a nonlocal quantity it is still a legitimate observable which, in principle, can be measured experimentally. Using the notation in Eq. (\ref{Schmidt-decomp}), and defining an operator ${\cal O}$ which acts on, say, the left part of the symmetrically cut system,
\begin{equation} \label{Observable}
{\cal O} \equiv |A_1\ra \la A_1|-|A_2\ra \la A_2|,
 \end{equation}
its expectation value in the ground state, $\la GS| {\cal O}|GS \ra=\lambda_1-\lambda_2$  is precisely the Schmidt gap. This means that having access
to only half of the system is enough to determine $\Delta_{\text{S}}$.  The operator ${\cal O}$ can be expanded in products of spin-1/2 operators, with the corresponding Schmidt gap encoded as a superposition of products of $n$-point spin correlations.  As this highly complex structure is not likely to be accessible with present-day experimental technology, its measurement offers a challenge for the future.

\subsection*{DISCUSSION}
In summary, we have shown that the Schmidt gap obtained from the entanglement spectrum provides a nonlocal order parameter for a quantum impurity system at criticality. A case study of the two-impurity Kondo model confirms that the Schmidt gap faithfully captures the scaling behavior by correctly predicting the critical exponent of the dynamically generated length scale at the quantum critical point.

Given our result that this length scale can be identified with an entanglement length, one may be tempted to speculate that its divergence at criticality $-$ together with the non-locality of the Schmidt gap order parameter $-$ is a smoking gun for the emergence of some kind of topological order as one enters the Kondo screened phase. However, this is not a viable proposition. The Kondo screened phase in the full two-impurity Kondo model with both spin and charge degrees of freedom is expected to be a smooth deformation of the single-impurity Kondo screened state and, as such, it simply enacts a renormalized Fermi liquid (with the quasiparticles experiencing potential scattering off the impurity site with a $\pi/2$ phase shift) \cite{Nozieres}. Intriguingly, here, it is rather the RKKY phase that carries some fingerprints of a topological phase: The degeneracy in the entanglement spectrum \cite{Pollmann} which makes the Schmidt gap vanish, and, as shown in Ref. \cite{SantoroGuiliani}, the presence of a gap in the local impurity spectral weight are both possible attributes of a symmetry-protected topological state. However, this resemblance is at best suggestive: As seen in Fig. \ref{fig2}(a), to the numerical accuracy allowed by DMRG, the degeneracy of the entanglement spectrum in the RKKY phase is only partial.

The physical interpretation of the Schmidt gap order parameter is indeed elusive. The non-vanishing of a local order parameter, as in the LGW paradigm, signals that a local symmetry is spontaneously broken. Similarly, one could expect that the non-vanishing of a nonlocal order parameter in a one-dimensional spin system $-$ as in a symmetry-protected  topological phase $-$ conveys that some nonlocal symmetry is broken \cite{Else}. Adapting this figure of thought to the present problem, one faces the challenge to find the hidden discrete symmetry that gets broken in the Kondo phase.  A candidate that may suggest itself is the emergent Z$_2$ Ising symmetry at the critical point which appears in the Bose-Ising decomposition of the model \cite{ALJ}. However, this symmetry is broken in both Kondo and RKKY phases and is therefore not immediately applicable for explaining the behavior of the Schmidt gap. Alternatively, could it be that the Schmidt gap plays the role of a nonlocal order parameter with no associated symmetry breaking, suggesting an entirely different kind of ``quantum order''? Maybe the search for answers to these questions will unveil new facets of critical quantum impurity physics.

\subsection*{METHODS}

\begin{center}{\bf{DMRG}}\end{center}
We use a DMRG approach \cite{white-DMRG} to target the ground state of our system. In this approach the Hilbert space of the system is truncated in such a way that the entanglement between any bipartition of the system is approximated by keeping the $M$ largest Schmidt numbers. In our code we keep $M=100$ Schmidt coefficients and use three sweeps in computing the ground state. \\

\begin{center}{\bf{Screening length from negativity}}\end{center}
To compute the entanglement screening length $L^*$ from the ground state of the system we compute the entanglement $-$ using negativity as an entanglement measure \cite{VidalWerner} $-$ between the two impurities (as a block $A$) and the rest of the system (as a block $B$) when $L$ spins in the proximity to each impurity are traced out from both the left and right chains, just as done in \cite{bayat-TIKM}. By varying $L$ the negativity decreases and when it becomes smaller than a threshold for a particular $L$ (with the threshold taken to be $0.01$ in this paper), we consider it  as the entanglement screening length $L^*$. \vspace{3mm} \\

\noindent {\bf Acknowledgements}\ \vspace{1mm} \\
\noindent It is a pleasure to thank S. Eggert and A. Mitchell for useful comments and suggestions. We acknowledge support from EPSRC grant EP/K004077/1 (nano-electronic based quantum technologies) (A.B.); Swedish Research Council grant no. 621-2011-3942 and STINT grant IG 2011-2028 (H.J.); and the ERC grant PACOMANEDIA (S.B.). P.S. thanks the Ministry of Science, Technology and Innovation of Brazil for financial support and CNPq for granting a "Bolsa de Produtividade em Pesquisa". \\ \\
\noindent{\bf Author contributions} \vspace{1mm} \\
\noindent A.B. carried out all numerical work and computations. A.B. and H.J. drafted the paper. All authors conceived the research, discussed the results, and contributed to the final version of the manuscript. \\ \\
\noindent {\bf{Additional information}} \vspace{1mm} \\
{\em Competing financial interests:} The authors declare no competing financial interests. \\
\noindent {\em Correspondence and requests for materials} should be addressed to A.B. (email: abolfazl.bayat@ucl.ac.uk).

\end{document}